\documentclass{pj}
\usepackage{graphicx}
\usepackage{amssymb}
\usepackage{amsthm}
\usepackage{bbm}
\usepackage{a4}
\usepackage[utf8]{inputenc}
\usepackage{enumerate}
\usepackage{epsf}
\usepackage{amsmath}
\usepackage{listings}
\usepackage{hyperref}

\begin{document}
\setcounter{page}{1}
\pjheader{Vol.\ x, y--z, 2004}

\title{Adaptive Row-grouped CSR format for storing of sparse matrices on GPU}

\author{Martin~Heller, Tomas~Oberhuber ({\tt tomas.oberhuber@fjfi.cvut.cz})}

\address{Department of mathematics, Faculty of Nuclear Sciences and Physical Engineering, Czech Technical University in Prague, Trojanova 13, Praha 2, 120 00, Czech Republic}  



\runningauthor{Oberhuber}
\tocauthor{T.~Oberhuber}

\begin{abstract}
We present new adaptive format for storing sparse matrices on GPU. We compare it with several other formats including CUSPARSE which is today probably the best choice for processing of sparse matrices on GPU in CUDA. Contrary to CUSPARSE which works with common CSR format, our new format requires conversion. However, multiplication of sparse-matrix and vector is significantly faster for many matrices. We demonstrate it on set of 1 600 matrices and we show for what types of matrices our format is profitable.
\end{abstract}

%

\lstset{language=C,
        basicstyle=\footnotesize,
        numbers=left,
        frame=single,
        firstnumber=auto}

\section{Introduction}
\label{sect:introduction}
GPUs are specialized devices offering high computational power as well as high memory throughput \cite{LeeKimChhuganiDeisherKimNguyenSatishSmelyanskiyChennupatyHammarlundSinghalDubey-2010,OberhuberSuzukiZabka-2011}. Their architecture is very similar to processor arrays \cite{Quinn-2003}. GPUs are efficient for algorithms exhibiting massively parallel and homogeneous computations. Even though most operations from linear algebra are not computationally intensive, corresponding algorithms can profit from high memory throughput reaching almost 200 GB/s. This has been demonstrated in many works dealing with dense matrices \cite{GaloppoGovindarajuHensonManocha-2005,NvidiaCublas-2010,VolkovDemel-2008}. Situation is more complicated in case of sparse matrices. They often have irregular pattern of non-zero elements and they significantly reduce available parallelism. It makes processing of the sparse matrices on GPU difficult but also challenging.

In this article we concentrate on the sparse-matrix vector multiplication (SpMV) since it is a key part of many iterative solvers for linear systems. 
The performance is influenced mainly by the format used to store the matrix. Formats for storing the sparse matrices often involves additional data, typically column indexes. On the vector architectures, the data must by aligned in the memory. CUDA developers speak about coalesced global memory accesses. Their importance for the efficiency of SpMV is discussed in \cite{BellGarland-2008,OberhuberSuzukiVacata-2011}. For this reason, artificial zeros must be often inserted. Moreover, with different number of non-zero elements in each row, the multiprocessors may be load unequally. Efficient format should address the following:

\begin{enumerate}
\item reduce amount of additional data while keep coalesced global memory accesses,
\item distribute the non-zero elements of the matrix evenly.
\end{enumerate}

\subsection{Contribution}
\label{sect:contribution}

We present new format for storing the sparse-matrices on GPU. It is optimized for matrix-vector multiplication. With some types of matrices, it is several times faster compared to today state-of-the-art formats like CUSPARSE, Hybrid (in CUSP library) \cite{BellGarland-2008}, Row-grouped CSR \cite{OberhuberSuzukiVacata-2011} and Sliced ELLPACK \cite{MonakovLokhmotovAvetisyan-2010}. We analyze on what type of matrices the new format outperforms the others and vice versa.

\subsection{Organization}
\label{sect:organisation}

The article is organized as follows. In Section \ref{sect:spmv-on-gpu}, we briefly summarize existing formats for storing sparse matrices on GPU. The new format is presented in Section \ref{sect:arg-csr-format} together with description of conversion from CSR format. The implementation in CUDA with a source code of matrix-vector multiplication kernel is in Section \ref{sect:cuda-implementation}. Achieved results with short performance analysis are topics of the last Section \ref{sect:results}.

\section{Formats for sparse matrices on GPU}
\label{sect:spmv-on-gpu}

\begin{figure}[]
\center{
\includegraphics[width=10cm]{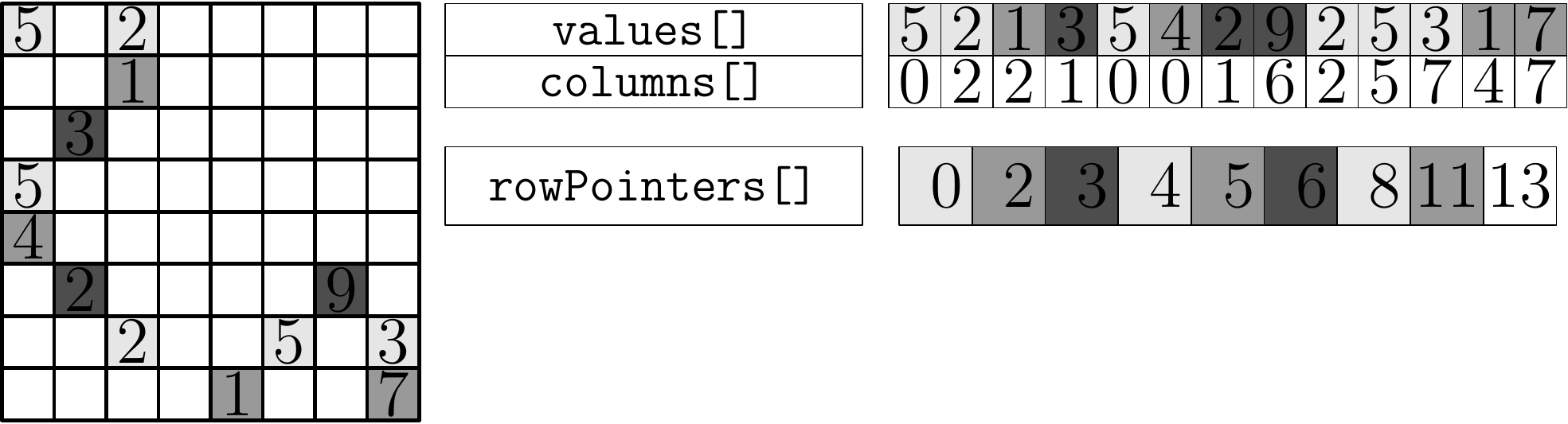}
}
\caption{CSR format}
\label{fig:csr}
\end{figure}

CSR (Compressed Sparse Rows) format (Figure \ref{fig:csr}) is standard and most popular format for storing sparse matrices \cite{Saad-2003}. It consists of arrays {\tt values} storing rowisely the non-zero elements of the matrix, {\tt columns} storing column index of each non-zero element and {\tt rowPointers} containing offset of each matrix row in arrays {\tt values} and {\tt columns}. This format is easy to implement. Storing data in arrays improves efficiency of data transfer.

\begin{figure}[]
\center{
\includegraphics[width=6cm]{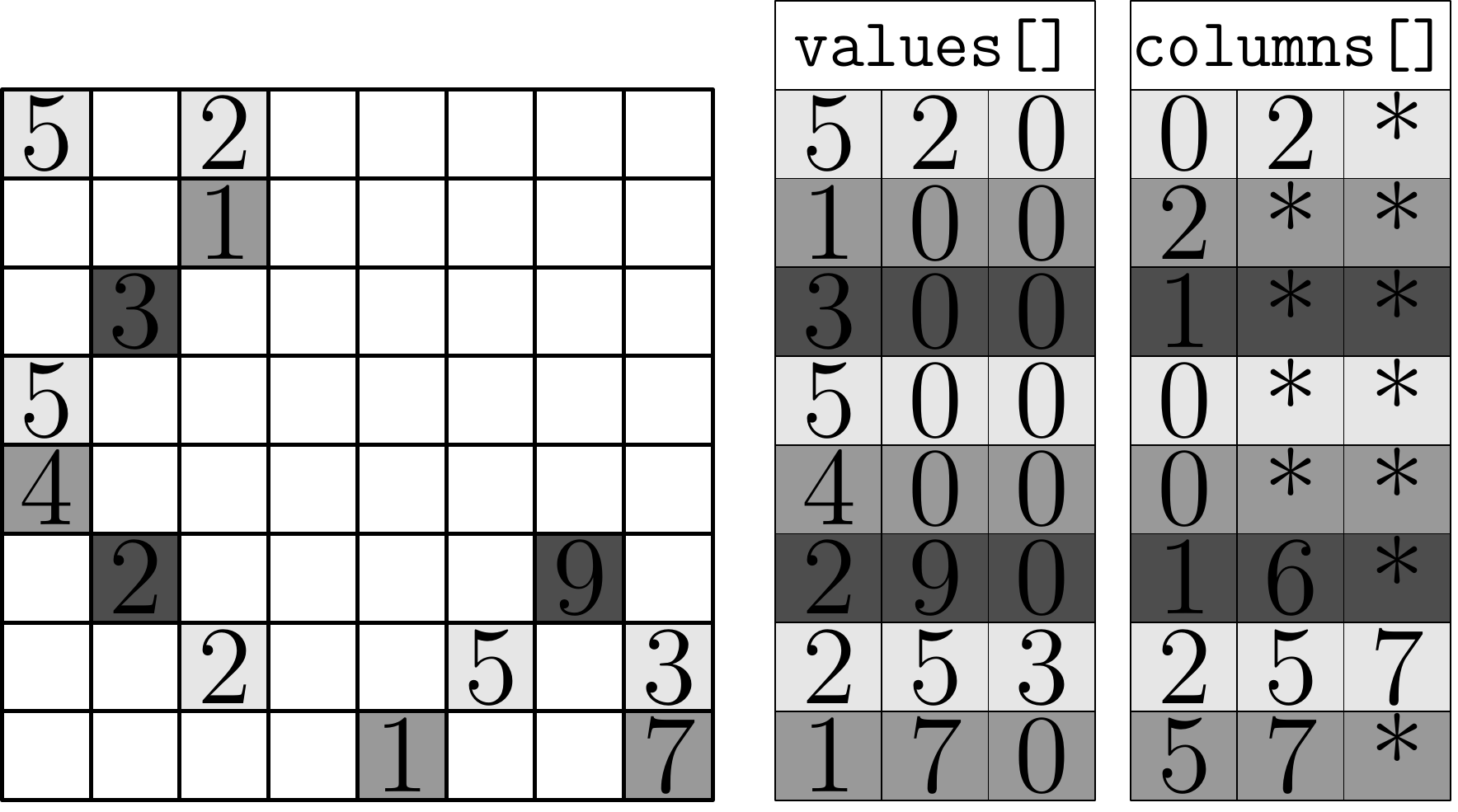}
}
\caption{ELLPACK format}
\label{fig:ellpack}
\end{figure}

Works by Bell and Garland \cite{BellGarland-2008} or Bautois et al. \cite{BuatoisCaumonLevy-2009} showed that this format is inefficient on GPU for matrix-vector multiplication. Better formats are based on the ELLPACK format (Figure \ref{fig:ellpack}) by Monakov and Avetisian \cite{MonakovLokhmotovAvetisyan-2010} or our similar format studied in \cite{OberhuberSuzukiVacata-2011}. Formats based on ELLPACK require homogeneous distribution of non-zero elements in rows. If the number of non-zero elements in each row is very different, the ELLPACK format loses efficiency. Bell and Garland \cite{BellGarland-2008} proposed Hybrid format which is part of CUSP library. It has also achieved great popularity. Recently CUSPARSE \cite{Nvidia-2012} showed that even pure CSR format can be implemented efficiently on GPU. Nevertheless, tests on large sets of sparse matrices like those in \cite{OberhuberSuzukiVacata-2011} show that there are a lot of matrices for which common CPU performs better. There is still great potential to improve formats for sparse matrices on GPU.
The ELLPACK format, as depicted on the Figure \ref{fig:ellpack}, allocates the same number of elements for each matrix row. If there are less non-zero elements on some row, artificial zeros are added to align the data. This increases memory requirements and slow down matrix-vector multiplication because more data must be transferred. On GPU, usually one thread is mapped to one row. Arrays {\tt values} and {\tt columns} are stored in global memory and so the accesses to these arrays must coalesced (see \cite{BellGarland-2008,OberhuberSuzukiVacata-2011}). This is reason why these arrays are stored columnwise instead of rowise. Since the coalesced memory accesses must be fulfilled only for threads in one warp we may split the matrix into slices of rows processed by the same warp. The slices are stored separately. If there is a row having significantly more non-zero elements than the others, only one slice is affected. This modification, introduced independently in \cite{MonakovLokhmotovAvetisyan-2010,OberhuberSuzukiVacata-2011}, reduces number of artificial zeros required by original ELLPACK format. However, for some matrices these formats still generate too many artificial zeros and matrix-vector multiplication can be several orders slower compared to CSR format on CPU.

The next step is to split long rows (i.e. with a lot of non-zero elements) and process them by more threads. The matrix is again divided into groups of rows. We introduce {\it chunks} of non-zero elements. Each chunk in the same group has the same size. One row can be splitted into more chunks (but one chunk cannot cross boundary of one row). Chunks are stored in the same way as matrix rows in the ELLPACK format (also stored columnwise to get the coalesced memory accesses). When performing the matrix-vector multiplication, we map one thread to each chunk. First the chunks are processed which results to array of partial sums. If some row consists of more chunks, its partial sums must be summed to get the final result. In the following text, we explain this format in more details.

\section{New Adaptive Row-grouped CSR format}
\label{sect:arg-csr-format}
\begin{figure}[]
\center{
\includegraphics[width=12cm]{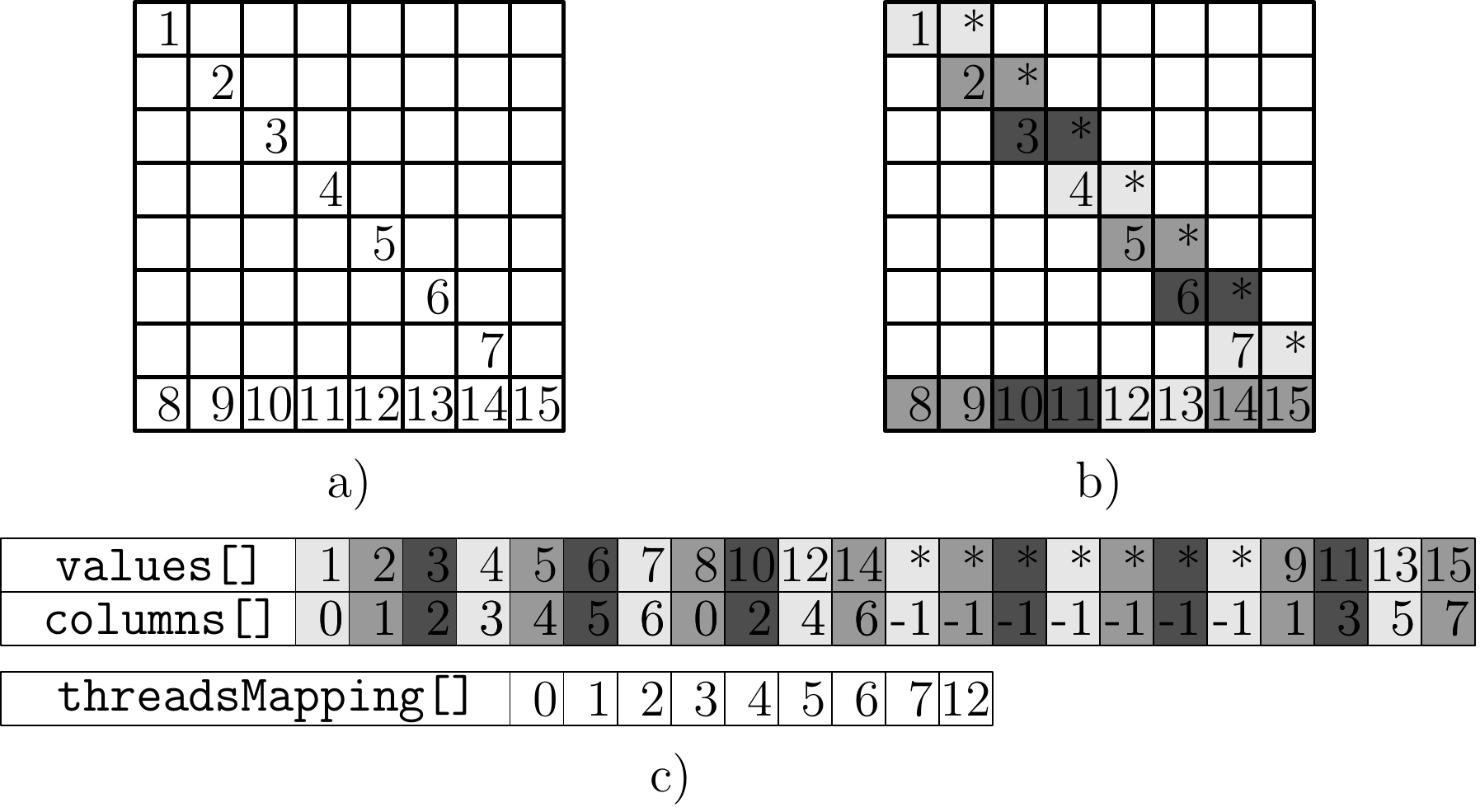}
}
\caption{Example demonstrating the Adaptive Row-grouped CSR format: a) the original matrix, b) mapping of chunks to matrix rows -- each chunk is depicted in different greyscale, c) arrays with the non-zero elements ({\tt values}), column indexes ({\tt columns}) and mapping of threads to rows ({\tt threadsMapping}).}
\label{fig:argcsr-1}
\end{figure}
Figure \ref{fig:argcsr-1} shows an example of matrix for which the ELLPACK format is inefficient. All rows except the last one have only one non-zero element and the last row is full. Assume that we map 12 threads to this matrix. We first assign one thread to each row. The thread mapped to the last row must process 8 elements. Therefore the chunk size is 8 and all other threads must process 8 elements as well because they belong to the same warp. We would have to allocate 49 artificial zeros to align the data. Since we have 4 threads left we may use them to diminish the chunk size by mapping them to the last row. As depicted in the Figure \ref{fig:argcsr-1}, if there are 4 threads assigned to the last row, the chunk size is only 2. One thread remains free and we need to allocate only 7 artificial zeros. To ensure the coalesced global memory accesses we store the chunks in the same way as rows are stored in Sliced ELLPACK format \cite{MonakovLokhmotovAvetisyan-2010} or Row-grouped CSR format \cite{OberhuberSuzukiVacata-2011}. In other words, if chunks were rows of some matrix, the matrix would be stored columnwise.

Large matrices are splitted into small groups of rows. Each group is defined by the following structure (Listing \ref{listing:group-info}):
\begin{lstlisting}[caption={Structure defining group of rows},label={listing:group-info}]
struct argcsrGroupInfo
{
   int firstRow;
   int size;
   int offset;
   int chunkSize;
};
\end{lstlisting}
Parameter {\tt firstRow} is the first matrix row and {\tt size} is number of rows in the group i.e. the group contains rows indexed from {\tt firstRow} to {\tt firstRow + size - 1}. Parameter {\tt offset} points to the first element of the group in arrays {\tt values} and {\tt columns} and {\tt chunkSize} describes number of elements in chunk. It is constant in each group but it can be different for different groups. The number of chunks is the same for each group and equals CUDA block size. For each group there are {\tt chunkSize * blockDim.x} elements in arrays {\tt values} and {\tt columns} starting at position {\tt offset}. 

Let us now comment a conversion from the CSR format. For good performance, it is necessary that all groups have approximately the same number of non-zero elements. Otherwise the load balance of multiprocessors would be unequal. One parameter, entering the converting algorithm is {\tt desiredChunkSize}. We allocate new group, read the matrix row by row and compute number of non-zero elements. Once it is larger than {\tt desiredChunkSize * blockDim.x} or there would be more rows than {\tt blockDim.x} in the current group, we close it and allocate new one. The groups are defined by array of structures {\tt argcsrGroupInfo}. In the next step of the conversion, we compute the chunk size in each group. This can be done in parallel. We start by mapping one thread to each row of the group. The chunk size would be now equal to the number of non-zero elements in row with greatest filling. We define {\it the chunk filling} as number of non-zero elements in given chunk. If there are some threads left, we always find row with greatest chunk filling and map one more thread to it. When all threads are distributed to the rows we can compute the final chunk size. We also need to store mapping of threads to rows. For this we allocate array {\tt globalThreadsMapping}. It contains number of threads mapped to each row of matrix. Then we perform exclusive prefix-sum on this array separately for each group.  The last step is filling the arrays {\tt values} and {\tt columns} by data belonging to particular chunks. Since we know the chunk size in each group we know how many elements will be inserted by each group. This phase thus can be done in parallel as well. We read the data chunkwise from CSR format and copy them in appropriate order to the mentioned arrays.

In the next section we explain matrix-vector multiplication.

\section{Implementation in CUDA}
\label{sect:cuda-implementation}
 
For better understanding, we show the source code of the kernel in CUDA -- see Listing \ref{listing:spmv-kernel}.

\begin{lstlisting}[caption={Kernel for matrix-vector multiplication},label={listing:spmv-kernel}]
template< class Real >
__global__ void spmvKernel( 
   Real* target,
   Real* vect,
   Real* values,
   int* columns,
   argcsrGroupInfo* globalGroupInfo,
   int* globalThreadsMapping )
{
   extern __shared__ int sdata[];
   const int* globalGroupInfoPointer = 
      reinterpret_cast< const int* >( globalGroupInfo );
   argcsrGroupInfo* groupInfo = 
      reinterpret_cast< argcsrGroupInfo* >( &sdata[ 0 ] );
   int* threadsMapping = 
      reinterpret_cast< int* >( &sdata[ 4 ] );
   Real* partialSums = 
      reinterpret_cast< Real* >( &sdata[ 4 + blockDim. x] );

   int bId = blockIdx.x;
   
   /****
    * Fetch the group info from the global memory
    */
   if( threadIdx.x < 4 )
      sdata[ threadIdx.x ] = 
         globalGroupInfoPointer[ 4 * bId + threadIdx.x ];
   __syncthreads();

   /****
    * Fetch mapping of threads to rows.
    */
   if( threadIdx. x < groupInfo -> size )
      threadsMapping[ threadIdx. x ] = 
         globalThreadsMapping[ groupInfo -> firstRow + threadIdx. x ];

   /****
    * Each thread computes partial sum in its chunk
    */
   Real sum = 0;
   int threadOffset = groupInfo -> offset + threadIdx. x;
   for( int i = 0; i < groupInfo -> chunkSize; i ++ )
   {
      const int column = columns[ threadOffset ];
      if( column != -1 )
         sum += values[ threadOffset ] * vect[ column ];
      else
         break;
      threadOffset += blockDim. x;
   }
   partialSums[ threadIdx. x ] = sum;
   __syncthreads();


   /****
    * Sum the partial sums in each row
    */
   if( threadIdx. x < groupInfo -> size )
   {
      sum = 0;
      int begin( 0 );
      const int row = groupInfo -> firstRow + threadIdx. x;
      if( threadIdx. x > 0 )
         begin = threadsMapping[ threadIdx. x - 1];
      int end = threadsMapping[ threadIdx. x ];
      for( int i = begin; i < end; i++ )
         sum += partialSums[ i ];
      target[ row ] = sum;
   }
}
\end{lstlisting}

The kernel first fetch data with reuse to fast shared memory. We allocate shared memory (lines 10--18) for one {\tt argcsrGroupInfo} structure, array {\tt threadsMapping} keeping track of thread indexes mapped to rows of the group and array {\tt partialSums} meaning of which is explained later. Then we fetch the group info structure. To achieve coalesced memory access we employ four threads to this task (lines 25--28). Thread synchronization is important here. Next, we may fetch array with mapping of threads to rows. The next part is independent on this array and therefore synchronization is not necessary. Each thread takes its own chunk and perform multiplication of this part of matrix data with given part of input vector. Result of this is partial sum. We remind that we mark artificial zeros by column index {\tt -1}. Once a thread reaches this column index it exits the loop on lines 42--50. If it happens in whole warp, it exits too and it omits the rest of artificial zeros. The last step is summing of the partial sums. There is data dependency with the previous part and so thread synchronization on the line 52 is necessary.

\section{Results}
\label{sect:results}

The results, we present in this section, were obtained on a system equipped with CPU AMD Phenom II X6 1100T with 16 GB DDR3 RAM and GPU Nvidia Tesla C2070 having 6 GB GDDR5 with memory bandwidth 144 GB/s (ECC was turned off). All tests were done in double precision and sequentially on CPU. Testing matrices were fetched from matrix databases \cite{BaiDayDemmelDongarra-1996,DavisHu-1994}. The testing set contained almost 1 600 sparse square matrices.

 The best performance was achieved with 128 threads in block. Test show that the parameter {\tt desiredChunkSize} can have strong impact on the efficiency. Simple rule might be: the more regular the matrix is (in sense that there are almost the same non-zero elements in each row), the larger the desired chunk size should be. With desired chunk size 32 we have achieved the highest performance almost 18 GFLOPS with matrices {\tt Schenk\_AFE} originating in structural problems. With {\tt desiredChunkSize} set to one, the performance dropped to 11 GFLOPS for this matrix. On the other hand, with the matrix {\tt rajat23}, the performance was six times higher with {\tt desiredChunkSize} 1 (5.1 GFLOPS and speed-up 11 compared to CSR on CPU) than with {\tt desiredChunkSize} 32 (0.81 GFLOPS). In the rest of this section we present results obtained with the desired chunk size set to 1 because it seems to be more robust setting.

\begin{figure}[]
\center{
\includegraphics[width=9cm,angle=-90]{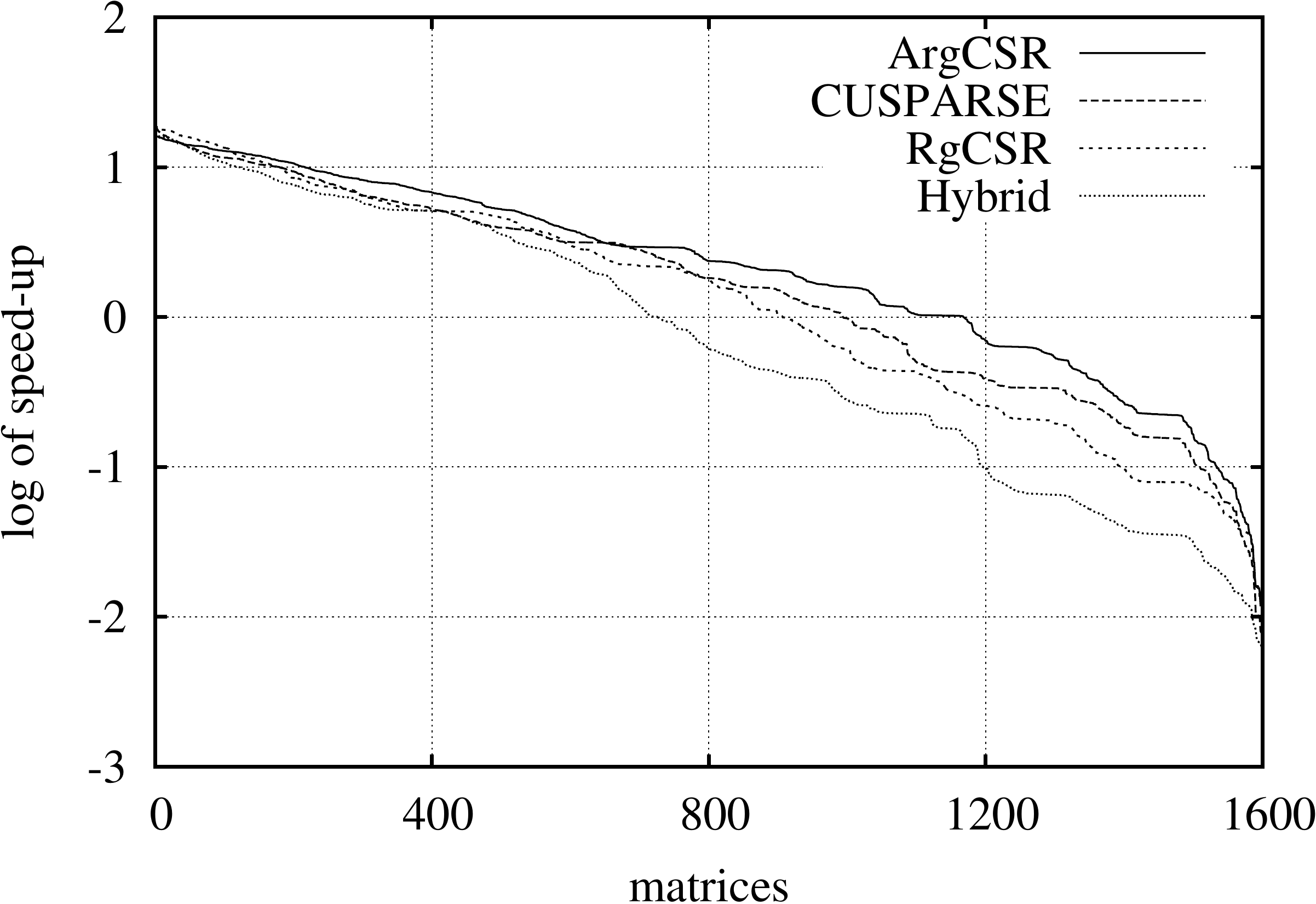}
}
\caption{Comparison with CSR format on CPU. Vertical axis shows logarithmic speed-up, horizontal axis shows on how many matrices was at least given speed-up achieved.}
\label{fig:csr-comparison}
\end{figure}

 Figure \ref{fig:csr-comparison} shows speed-up of tested formats compared to CSR format on CPU. The vertical axis shows the speed-up in logarithmic scale and the horizontal axis shows on how many matrices the formats attain given speed-up. There is one curve for each format and the slower it decreases the better the format is. Our test shows that the Hybrid format, Row-grouped CSR format, CUSPARSE library and the new format are is faster for 726, 907, 994 and 1168 matrices of 1600 respectively. This figure also shows that there are few matrices for which CPU is two orders of magnitude faster. They are mainly small matrices having tens or hundreds of unknowns as our previous test in \cite{OberhuberSuzukiVacata-2011} show.

\begin{figure}[]
\center{
\includegraphics[width=9cm,angle=-90]{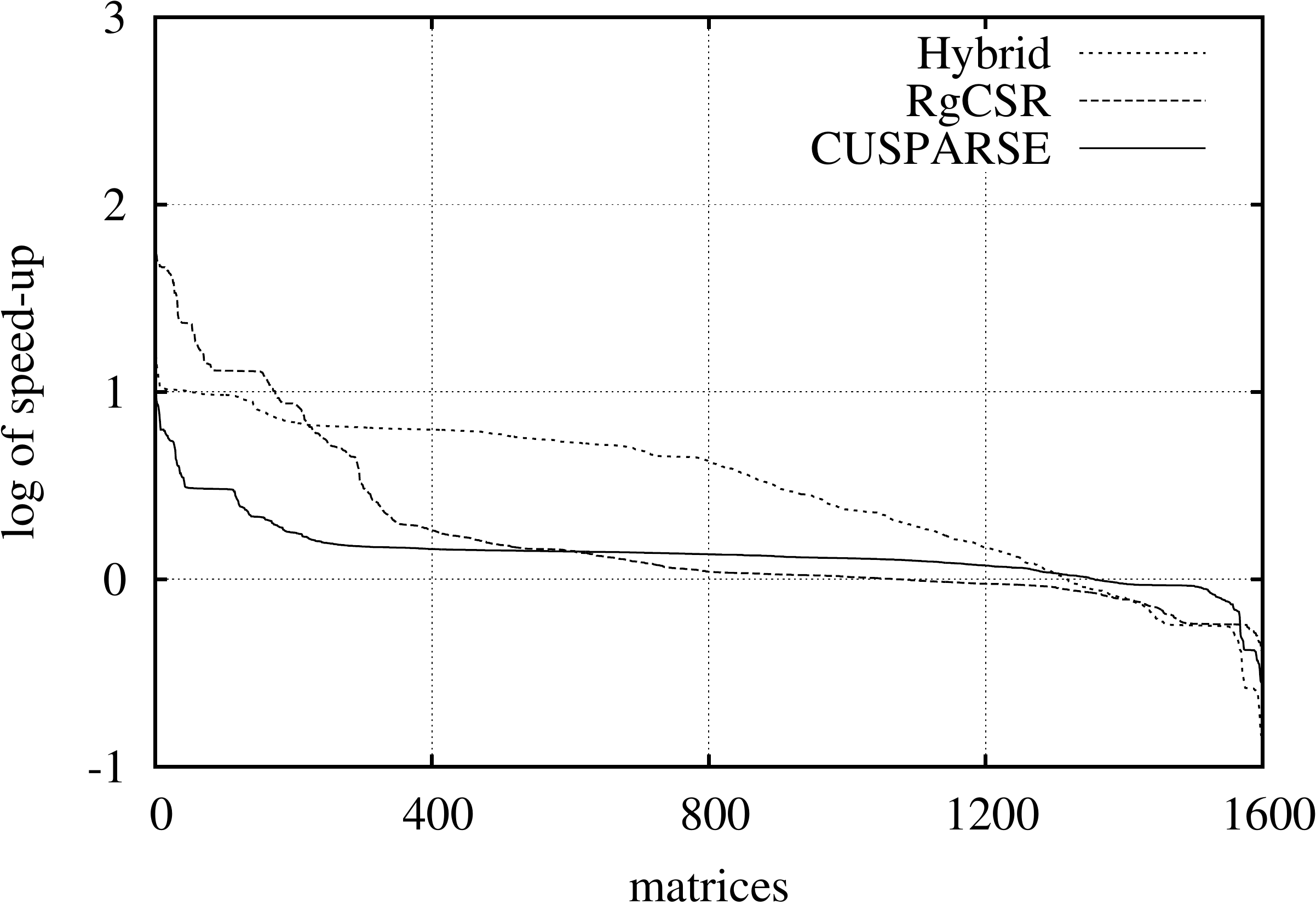}
}
\caption{Evaluation of the speed-up of the new format compared to other formats. Vertical axis shows logarithmic speed-up of the new format, horizontal axis shows on how many matrices was at least given speed-up achieved.}
\label{fig:argcsr-comparison}
\end{figure}

Figure \ref{fig:argcsr-comparison} shows speed-up of the new format compared to the others. The vertical axis shows speed-up in logarithmic scale while the horizontal number of matrices for which the new format attains given speed-up compared to other formats. The higher the curve is, the better the new format is. Our test show that the the Hybrid format is slower on 1318 matrices, Row-grouped CSR on 1072 matrices and CUSPARSE on 1358 matrices. 

Both Figures \ref{fig:csr-comparison} and \ref{fig:argcsr-comparison} demonstrate that the performance of sparse-matrix and vector multiplication on GPU is very variable. If high performance is the top priority, one should test more formats and choose the best one. For example the CUSPARSE library is almost 4 times faster than the new format on {\tt TSOPF} and {\tt case39} matrix sets from \cite{DavisHu-1994}. Both types of matrices model "Transient stability-constrained optimal power flow" and usually the Hybrid format outperforms the others in these cases. On the other hand, the new format is ten times faster than CUSPARSE (and at the same time 5-8 times faster than CSR on CPU) for matrices {\tt raj}, {\tt rajat}, {\tt GHS\_indef}, {\tt IBM\_EDA}. These matrices come from circuit simulations or optimizations problems. Original Row-grouped CSR format (or similar Sliced ELLPACK) is almost twice faster for {\tt norris/torso2} and {\tt t2d\_q} matrices originating in finite difference methods. The mentioned matrices are all difficult to visualize in this paper because of very large dimensions and we refer reader to \cite{DavisHu-1994} for more details.

\section*{Acknowledgment}
The source code of the format is freely available as a part of the Template Numerical Library (TNL) at 
\begin{center}
\href{http://geraldine.fjfi.cvut.cz/~oberhuber/doku-wiki-tnl}{http://geraldine.fjfi.cvut.cz/\textasciitilde  oberhuber/doku-wiki-tnl}.
\end{center}

This work was partially supported by the Jind\v{r}ich Ne\v{c}as Center for Mathematical Modelling, Research center of the Ministry of Education of the Czech Republic LC06052, Research Direction Project of the Ministry of Education of the Czech Republic No. MSM6840770010, and Advanced Supercomputing Methods for Implementation of Mathematical Models, project of the Student Grant Agency of the Czech Technical University in Prague No. SGS11/161/OHK4/3T/14, 2011-13.








\end{document}